\documentstyle{article}
\begin{document}
\large

\begin{center}
\baselineskip 10mm
{\huge Generalized gradient expansions in \\
quantum transport equations}\\
\vspace{5mm}
{\Large Petr Kr\'al}\\
\vspace{5mm}
\baselineskip 6mm
{\it Institute of Physics, Academy of Sciences,
Na Slovance 2,\\ 180 40 Praha 8, Czech Republic} \\
\end{center}
\vspace{5mm}
\begin{abstract}
Gradient expansions in quantum transport equations of a Kadanoff--Baym
form have been reexamined. We have realized that in a consistent
approach the expansion should be performed also inside of the 
self--energy in the scattering integrals of these equations. In the
first perturbation order this internal expansion gives new correction 
terms to the generalized Boltzman equation. These correction terms 
are found here for several typical systems.  Possible corrections to 
the theory of a linear response to weak electric fields are also discussed.
\end{abstract}
\newpage
\baselineskip 6mm

\newpage
%%%%%%%%%%%%%%%%%%%%%%%%%%%%%%%%%%%%%%%%%%%%%%%%%%%%%%%%%%%%%%%%%%%%%%%%%%%%%%%
\section{Introduction}
%%%%%%%%%%%%%%%%%%%%%%%%%%%%%%%%%%%%%%%%%%%%%%%%%%%%%%%%%%%%%%%%%%%%%%%%%%%%%%%
Time dependent transport phenomena in quantum many--body systems can be 
described by the nonequilibrium Green's function formalism (NGF) of 
Kadanoff and Baym \cite{KaBa} or Keldysh \cite{Keldysh}. The
Kadanoff--Baym transport equations for nonequilibrium correlation functions 
can be obtained by analytic continuation to real times 
\cite{Langreth} of the Dyson equation for Matsubara Green's functions 
in purely complex times \cite{Abrikosov,Mahanb}. The differential 
form of these equations has been applied in many systems 
\cite{LangrethMag,Danielewicz,Schafer,Hennerberger}. Usually it is 
necessary to approximate the equations on several levels \cite{KaBa}, 
although in some systems the equations can be directly solved by 
powerful numerics \cite{LangrethSur}. The integral form of the 
Kadanoff--Baym equations was less exploited, because approximations in
a time domain are not so familiar here \cite{Kalvova}. Both these approaches 
have been also used to develop a linear response theory for quantum 
systems in weak $dc$ and $ac$--electric fields. Here the integral 
version \cite{Kral} seems to be more direct than the older differential 
versions \cite{Hansch,Wu}.

All functions in the Kadanoff--Baym equations depend separately on two 
time and two space arguments $(r_1,r_2;t_1,t_2)$. This two--argument 
structure, which result in nonequilibrium many--body
systems with time or space nonlocal scattering, is the main obstacle in 
solving the transport equations. An approximate one--argument form of
the equations can be obtained in systems with not very strong
interactions \cite{Lipavsky}. Most older transport methods have a 
one--argument structure, because they usually implicitly consider
the presence of weak interactions.

A different simplification can be obtained in all types of systems
if external excitation fields vary slowly in time and space. Then it 
is useful to subtract two equivalent sets of the 
Kadanoff--Baym equations, differentiated over the first $(r_1,t_1)$ 
and second variables $(r_2,t_2)$, and perform the so called gradient 
expansion \cite{KaBa} in the new equations transformed to the 
center--of--mass system (CMS) $\xi=(x,X)= (r,t;R,T) 
=\left(r_1-r_2,t_1-t_2 ;\frac{r_1+r_2}{2},\frac{t_1+t_2}{2} \right)$. 
In the transformed equations various terms have the second arguments 
with big variables $(R,T)$ shifted by different fractions of the 
small variables $(r^{'},t^{'})$. A Taylor expansion of these terms
in the small variables arround the common values of the big variables
is the gradient expansion. This expansion in powers of derivatives 
over $(R,T)$ multiplied by $(r^{'},t^{'})$ can be {\it unambiguously} and
{\it consistently} stopped at chosen perturbation orders. In this way
the two sets of CMS variables $(R,T)$ and $(r^{'},t^{'})$ can be step
by step decoupled. By this decoupling memory effects are cuted, so that
the nonequilibrium dynamics becomes quasi--local and the equations get
a quasi--equilibrium form. Far from equilibrium this approach evidently 
fails, because the nonlocal scattering, leading (in equilibrium) to
quasiparticles with a $(k,\omega)$--dependent self--energy, should be 
reflected in the nonlocality of the nonequilibrium dynamics.

In the zeroth order of the gradient expansion the transport equations 
get a local dynamics.  Stopping the gradient expansion in the first 
order gives corrections to the equations which partially restore 
memory effects peculiar to the nonlocal dynamics. These equations are 
called a generalized Boltzmann equation \cite{KaBa} (GBE). The GBE is 
not limited to weak interactions, but it can only describe slow 
dynamics close to equilibrium. The weaker is the scattering the better 
is the description far from equilibrium.  When the scattering processes 
can be considered extremally weak, then the nonequilibrium correlation 
functions in the GBE can be substituted by a delta--like spectral 
function multiplied with a distribution function for momenta \cite{KaBa}. 
As a result the GBE reduces to the Boltzmann equation (BE), which differs 
from its classical counterpart only by the degeneracy of the described
gases. It is interesting that the integral version of the quantum 
transport equations \cite{Jauho} approximated by low orders of the 
gradient expansion do not give the BE. It would be also good to 
mention that the gradient expansions in quantum transport equations 
are analogous to expansions in the classical Enskog's equation or the 
more general BBGKY equations \cite{Ferziger}. 

The gradient corrections in the GBE described in the past \cite{KaBa}
do not fully reflect the character of scattering processes, because 
the self--energy in the scattering integrals of the transport equations 
is considered as a structure--less entity. We have realized 
\cite{ThesKral} that consistently performed gradient expansion should 
include also expansion of the self--energy {\it itself}, as soon as 
the self--energy includes additional scattering events separated by
internal vertices \cite{Abrikosov}.  In the first order this 
{\it internal} expansion gives new correction terms to the GBE, 
which are determined by the character of many--body scattering processes.

In this work phenomenological rules are presented, which allow to 
perform these internal gradient expansions. The rules are applied 
to consistently derive the GBE with all correction terms. These
correction terms are found in three examples of a self--energy: 
the averaged $T$--matrix approximation for a self--energy in electron 
scattering on local potentials, the local $T$--matrix approximation 
and the shielded potential approximation for a self--energy of interacting 
spinless Fermions. The importance of the internal gradient corrections 
for a linear response to weak electric fields is also briefly discussed.

\newpage
%%%%%%%%%%%%%%%%%%%%%%%%%%%%%%%%%%%%%%%%%%%%%%%%%%%%%%%%%%%%%%%%%%%%%%%%%%%%%%
\section{Gradient expansions in quantum transport equations}
%%%%%%%%%%%%%%%%%%%%%%%%%%%%%%%%%%%%%%%%%%%%%%%%%%%%%%%%%%%%%%%%%%%%%%%%%%%%%%%
In slowly changing fields the two differential forms of the Kadanoff--Baym 
equations in (\ref{KBE}) can be subtracted and the gradient expansion can
be performed in the resulting equations. The right sides of the 
subtracted equations include scattering terms of the form
\begin{equation}
\Sigma^{\alpha}(x_1,\bar{x}_3)\ G^{\beta}(\bar{x}_3,x_2)\ , \ \
G^{\alpha}(x_1,\bar{x}_3)\ \Sigma^{\beta}(\bar{x}_3,x_2)\ , \ \ \ 
x_i=(r_i,t_i)\ , 
\label{SCAT}
\end{equation}
where the analytical structure of the self--energy $\Sigma$ and the 
Green's function $G$ in (\ref{SCAT}) is determined by the index 
$\alpha,\ \beta=r,\ a,\ <,\ >$ (see the Appendix A). 

%==============================================================================
\subsection{External expansions}
%==============================================================================
Consider as an example the gradient expansion in the term 
$\Sigma^{\alpha}(x_1,\bar{x}_3)\ G^{\beta}(\bar{x}_3,x_2)$ from
(\ref{SCAT}), where the many--body structure of $\Sigma^{\alpha}$ 
is neglected \cite{KaBa}.  The expansion results in the CMS coordinates 
$\chi=(x,X)=(r,t;R,T)$ as follows
\vspace{1mm}
$$
\int dx_3\ 
\Sigma^{\alpha}(x_1,x_3)\ G^{\beta}(x_3,x_2)
= \int dx_3\ 
\Sigma^{\alpha}\left(x_1-x_3,\frac{x_1+x_3}{2}\right)\
G^{\beta}\left(x_3-x_2,\frac{x_3+x_2}{2}\right)
 \vspace{1mm}
$$
$$
= \int
d\bar{x}\ \Sigma^{\alpha}\left(\bar{x},X+\frac{x-\bar{x}}{2}\right)\
G^{\beta}\left(x-\bar{x},X-\frac{\bar{x}}{2}\right)
 \vspace{1mm}
$$
$$
=\int
d\bar{x}\ \Sigma^{\alpha}(\bar{x},X)\ G^{\beta}(x-\bar{x},X)
+\int d\bar{x}\
\frac{\partial \Sigma^{\alpha}(\bar{x},X)}{\partial X} 
\left( \frac{x-\bar{x}}{2} \right) 
G^{\beta}(x-\bar{x},X)
 \vspace{1mm}
$$
\begin{equation}
+ \int d\bar{x} \
\Sigma^{\alpha}(\bar{x},X)\ 
\frac{\partial G^{\beta}(x-\bar{x},X)}{\partial X}\ 
\left( -\frac{\bar{x}}{2} \right) +... \ ,\
 \vspace{1mm}
\label{SLOW1}
\end{equation}
The new coordinates are equal to $x=x_1-x_2\ ,\ \bar{x}=x_1-x_3\ ,\
X=\frac{x_1+x_2}{2}$.  A Fourier transform of the last expression 
in (\ref{SLOW1}) from the coordinates $x=(r,t)$ to $q=(k,\omega)$ gives
the following expansion in a series of Poisson brackets \cite{Hennerberger} 
$$
\exp \left( \frac{i}{2}\ D(\xi,\xi') \right)\ 
\Sigma^{\alpha}(\xi)\ G^{\beta}(\xi')
\equiv \Sigma^{\alpha}(\xi)\ G^{\beta}(\xi)
 \vspace{2mm}
$$
$$
+ \frac{i}{2}
\left( \frac{\partial \Sigma^{\alpha}(\xi)}{\partial R}
\frac{\partial G^{\beta}(\xi)}{\partial k}
     - \frac{\partial \Sigma^{\alpha}(\xi)}{\partial k}
     \frac{\partial G^{\beta}(\xi)}{\partial R} 
     \right.  \left.
     -\frac{\partial \Sigma^{\alpha}(\xi)}{\partial T} 
     \frac{\partial G^{\beta}(\xi)}{\partial \omega}
     +\frac{\partial \Sigma^{\alpha}(\xi)}{\partial \omega} 
     \frac{\partial G^{\beta}(\xi)}{\partial T} 
     \right)+...
 \vspace{2mm}
$$
\begin{equation}
= \Sigma^{\alpha}(\xi)\ G^{\beta}(\xi) + \frac{i}{2}\ 
[ \Sigma^{\alpha}(\xi), G^{\beta}(\xi) ]+ ... \ ,\ \ \ 
\xi=(q,X)=(k,\omega;R,T)\ .
 \vspace{2mm}
 \vspace{2mm}
\label{SLOW2}
\end{equation}
The gradient expansion of the scattering term $\Sigma^{\alpha}
(x_1,\bar{x}_3)\ G^{\beta}(\bar{x}_3,x_2)$ has been obtained in 
(\ref{SLOW2}) without taking into account the many--body structure of the 
self--energy $\Sigma^{\alpha}$.  Therefore this expansion can be called 
{\it external}.  The first order of this expansion includes only the 
first Poisson bracket in (\ref{SLOW2}). Similarly can be performed the 
gradient expansion in driving terms of the quantum transport equations.

%==============================================================================
\subsection{Internal expansions}
%==============================================================================
In many systems the self--energy is approximated by Feynman diagrams 
formed by ladders or bubbles of Fermion and Boson Green's functions 
\cite{KaBa,Schafer,White}. This nontrivial functional of Green's 
functions, connected by internal integrations, can substitute the 
self--energy in any step of calculations. The internal structure 
of the functional was not taken into account in the gradient 
expansion (\ref{SLOW2}). Does it contribute by new terms in the 
gradient expansion? It is obvious that this question can be answered, 
if a systematic gradient expansion is performed in the expressions 
(\ref{SCAT}), where the self--energy is substituted by the functional 
of the Green's functions.

Consider for simplicity that the self--energy describes electron 
scattering on the localized potentials $V(r)=\sum_{r_i} V_0\ \delta(r-r_i)$.
Assume further that the value $V_0$ is of a moderate strength,
but the number the random coordinates $r_i$ is relatively small.
Then the suitable self--energy for this problem results by the averaged 
$T$--matrix approximation \cite{Economou} (ATA)
\vspace{2mm}
\begin{equation}
\Sigma(t_1,t_2)=c\ \{ V_0+ V_0^2\ G(t_1,t_2)
+ V_0^3\ G(t_1,\bar{t_3})\ G(\bar{t_3},t_2) + ... \} 
=c\ \theta(t_1,t_2)\ .
\label{STM}
\end{equation}
Here $c$ represents the weak concentration of the potentials $V_0$ and
$\theta$ is the local $T$--matrix. The Green's functions in (\ref{STM})
depends only on the time variables, because the space variables have been 
integrated out due to the local scattering.

Analytical continuation of the self--energy (\ref{STM}) gives the
propagator and correlation functions for the self--energy in the 
form \cite{KaBa} (see the Appendix B)
\vspace{2mm}
\begin{equation}
\Sigma^r(t_1,t_2)=c\ \{V_0 + V_0^2\ G^r(t_1,t_2)
+ V_0^3\ G^r(t_1,\bar{t_3})\ G^r(\bar{t_3},t_2) + ... \}
=c\ \theta^r(t_1,t_2)\ , 
\label{STMr}
\end{equation}
\begin{equation}
\Sigma^<(t_1,t_2) =c\
\theta^r(t_1,\bar{t_3})\ G^<(\bar{t_3},\bar{t_4})\ \theta^a(\bar{t_4},t_2)
=c\ \theta^<(t_1,t_2)\ .
\label{STM<}
\vspace{2mm}
\end{equation}
Each term from the expressions for $\Sigma^{r,<}$ in (\ref{STMr}--\ref{STM<})
is formed by several propagators or a correlation function with time 
arguments in a 'series' (see the Appendix B).

Let us study in details the gradient expansion of the scattering terms 
in (\ref{SCAT}) for the self--energy in (\ref{STMr}--\ref{STM<}). Assume 
for the beginning that the self--energy in the first expression from 
(\ref{SCAT}) is substituted by one of the terms from 
(\ref{STMr}--\ref{STM<}), which has two Green's functions $\Delta 
\Sigma(t_1,t_3) =c\ V_0^3\ G^{\alpha}(t_1,\bar{t}_2)\ G^{\beta}(\bar{t}_2,t_3)$
(no index is used at the contribution to the self--energy $\Delta \Sigma$
to show its analytical structure).  Then the expression 
$\Delta \Sigma(t_1,\bar{t}_2)\ G^{\gamma}(\bar{t}_2,t_3)$ can be transformed 
to the CMS coordinates as follows (we suppress the prefactor $c\ V_0^3$
and neglect the fact that the function $G^{\gamma}$ depends also on 
space variables)
$$
G^{\alpha}(t_1,\bar{t}_2)\ G^{\beta}(\bar{t}_2,\bar{t}_3)\ 
G^{\gamma}(\bar{t}_3,t_4)
 \vspace{2mm}
$$
$$
=G^{\alpha}\left(t_1-\bar{t}_2,\frac{t_1+\bar{t}_2}{2}\right)\ 
 G^{\beta}\left(\bar{t}_2-\bar{t}_3,\frac{\bar{t}_2+\bar{t}_3}{2}\right)\
 G^{\gamma}\left(\bar{t}_3-t_4,\frac{\bar{t}_3+t_4}{2}\right)
 \vspace{2mm}
$$
$$
=G^{\alpha}\left(\bar{\tau}_I,
  T+\frac{\tau-\bar{\tau}_I-\bar{\tau}_{II}}{2}
  +\frac{\bar{\tau}_{II}}{2}\right)\
 G^{\beta}\left(\bar{\tau}_{II},
 T+\frac{\tau-\bar{\tau}_I-\bar{\tau}_{II}}{2}-\frac{\bar{\tau}_I}{2}\right)\
 \vspace{2mm}
$$
\begin{equation}
\times 
 G^{\gamma}\left(\tau-\bar{\tau}_I-\bar{\tau}_{II},
 T-\frac{\bar{\tau}_I}{2}-\frac{\bar{\tau}_{II}}{2}\right)\ , \ \ \ \
 \alpha,\ \beta,\ \gamma=r,\ a,\ <,\ >\ ,
 \vspace{2mm}
\label{LIN0}
\end{equation}
where it holds $(\tau,T) \equiv (t_1-t_4, \frac{t_1+t_4}{2})$ and 
$\bar{\tau}_I=t_1-\bar{t}_2$, $\bar{\tau}_{II}=\bar{t}_2-\bar{t}_3$. In the
expression (\ref{LIN0}) a Taylor expansion in the $\tau$--coordinates 
can be performed around the $T$--coordinates as in (\ref{SLOW1}).
In a first order each of the Green's functions differentiated over $T$ 
has multiplicative coefficients formed by the $\tau$--coordinates of 
the remaining two Green's functions in a series. Therefore in this 
structure of arguments the Green's functions are equivalent from the 
point of view of gradient expansions.

After a Fourier transform over the small variables a symmetrical 
expression in the three functions $G^{\alpha,\beta,\gamma}$ 
with respect to the derivatives $\frac{\partial }{\partial X}\ 
\frac{\partial }{\partial q}$ can be obtained.  Therefore the complete 
gradient expansion up to the linear order can be shortly written as 
follows (the space variables have been included in $G^{\gamma}$)
$$
c\ V_0^3\ \left\{ G^{\alpha}(\omega,T)\ G^{\beta}(\omega,T)\ G^{\gamma}(\xi) 
+ \frac{i}{2}\ [G^{\alpha}(\omega,T)\ G^{\beta}(\omega,T)\ ,G^{\gamma}(\xi)]
\right.
 \vspace{2mm}
$$
\begin{equation}
\left.
+ \frac{i}{2}\ [G^{\alpha}(\omega,T),G^{\beta}(\omega,T)]\ 
G^{\gamma}(\xi) \right\}\ , \ \ \ \xi =(k,\omega;R,T)\ ,
 \vspace{2mm}
\label{LIN3}
\end{equation}
where the Poisson brackets have been used.
Since the space coordinates ($k,R$) are integrated out in the present 
self--energy, gradient expansions cannot be performed in these variables. 
But in a general case both pairs of coordinates contribute in the way
shown in (\ref{LIN3}). 

The second term in (\ref{LIN3}) resulted as in (\ref{SLOW2}) by the 
(external) gradient expansion of the expression $\Delta 
\Sigma(t_1,\bar{t}_2)\ G^{\gamma}(\bar{t}_2,t_3)$. The last term in 
(\ref{LIN3}) resulted by the gradient expansion of the internal structure 
of the self--energy contribution $\Delta \Sigma$.  Therefore this
expansion and the resulting term can be called {\it internal}. In (\ref{LIN3}) 
only contributions to the self--energy diagrams with two Green's 
functions were included.  Analogously can be performed the gradient 
expansion in terms with any number of Green's functions. This problem 
is solved in the next section in details.

The {\it internal} gradient corrections can be easily physically 
understood on the previous example, where scattering processes can be 
seen as many consequent events on the same center. If the system is 
excited by a time dependent field, then scattering conditions on the
center can change between these consequent scattering events. Corrections 
to these changed scattering conditions are represented by the internal 
correction terms.  For centers which are little smeared in space, internal 
gradient corrections result nonzero even in excitation by static fields.

\newpage
%%%%%%%%%%%%%%%%%%%%%%%%%%%%%%%%%%%%%%%%%%%%%%%%%%%%%%%%%%%%%%%%%%%%%%%%%%%%%%
\section{Gradient expansions in general}
%%%%%%%%%%%%%%%%%%%%%%%%%%%%%%%%%%%%%%%%%%%%%%%%%%%%%%%%%%%%%%%%%%%%%%%%%%%%%%%
A gradient expansion in quantum transport equations can be {\it unambiguously} 
represented by a series of Poisson brackets of an increasing order $n$.  
Physically this is a consistent expansion in powers of space (time) 
inhomogeneities, since the $n$-th order 
Poisson brackets could be appreciated by terms of the form $(k\sigma)^n$,
where $k$ is the inverse mean free path and $\sigma$ is the range 
of space inhomogeneity (analogous terms apply for the time
inhomogeneity). If $k\sigma \ll 1$ then the expansion can be stopped
in the first order ($n=1$), which is equivalent to inclusion of both 
{\it external} and {\it internal} corrections in the GBE (see also 
(\ref{LIN3}), (\ref{LIN4}) and (\ref{GD<})). 

Gradient expansions can be performed also in classical transport equations 
describing dense systems. Such systems have been firstly approximately 
studied by the Enskog's equation \cite{Ferziger}, which generalizes 
the Boltzman equation by taking finite volumes of scattering particles.
Since only binary collisions are included here, like in the BE, gradient 
expansions in this equation are analogous to the {\it external} 
expansions in quantum transport equations, giving the uncomplete quantum 
GBE \cite{KaBa}.
Later on classical dense systems have been described by the so called 
BBGKY hierarchy of kinetic equations \cite{Ferziger}, which can include 
multiple encounters of particles. Under some 
assumptions this set of equations can be reduced to a classical generalized
Boltzman equation \cite{Ferziger}, which has scattering integrals with 
a structure analogous to that in the quantum transport equations. Therefore
gradient expansions in this classical GBE are analogous to both the 
{\it external} and {\it internal} expansions leading to the complete
quantum GBE (in the classical GBE gradient expansions can be performed, 
while the quantum GBE is the result of the gradient expansions).

%==============================================================================
\subsection{General rules for gradient expansions}
%==============================================================================
We formulate a set of phenomenological rules for performing {\it
external} and {\it internal} gradient expansions in terms like in
(\ref{SCAT}), appearing in quantum transport equations.  From now the 
gradient expansions are stopped after the first perturbation order (higher 
order contributions can be found analogously). The following 
qualitatively different points specify and summarize the necessary steps 
for gradient expansions in the concret terms: 
\begin{enumerate}
\item
Analytic continuation to real times of the term is performed, to get an
expression formed by propagators and correlation functions.
\item
Each of the self--energy functions $\Sigma^{r,a,<,>}$ is resolved 
into a functional
of full Fermion and Boson Green's functions $G^{r,a,<,>}$ and
undressed matrix elements.
\item
The whole term is transformed into CMS coordinates, and the big
coordinates $X$ are linearized in the conjugated small coordinates $x$.
After a Fourier transform over the coordinates $x$, the linearization 
prefactors $x$ become derivatives $\partial_q$, which produce 
a series of new terms. In each of these new terms just two 
derivatives appear (over $X$ and $q$). The expansion should be performed 
on a hierarchical structure of levels, going more and more inside to
the structure of  $\Sigma$ (the vertices are the landmarks).  Corrections 
for higher levels are done on lower levels.
All objects, which depend on some of the coordinates, must be differentiated.
The lowest objects are the full Green's functions and undressed coupling 
matrix elements.
\item
'Parallel' objects with the same coordinates $(x_i,x_j)$, but the order
of $x_{i,j}$, can be considered in the derivatives as a single 
differentiated object. At the lower level some of the objects with the 
coordinates $(x_i,x_j)$ can have still other internal coordinates, which 
can give further gradient corrections.
\item
'Serial' objects with arguments like in (\ref{LIN0}) are
differentiated in such a way that differentiation of one object over 
a big coordinate $X$ is accompanied by differentiation of the other 
objects in a series (one by one) over the conjugated small coordinates
$q$.  Sign prefactors depend on the sequence of objects.
Realization of this rule for terms with many Green's functions in a
series can be done by a second functional derivative of the term over 
the objects in the series. This derivative is multiplied by a Poisson 
bracket of the two differentiating objects in the functional derivative.
\end{enumerate}

In the previous section these rules have been already implicitly applied 
on a simple algebraic term from a self--energy (see (\ref{LIN3})).
 Analogously can be 
dealed other such algebraic terms or complex recursive terms. Before 
we come to these expansions let us still present one theorem.
Application of the above rules to scattering terms from transport 
equations, which include a self--energy as in (\ref{LIN3}), gives the
following formal expansion (stopped in the first perturbation order) 
\begin{equation}
G^{\alpha}(\xi)\ \Sigma(\xi) + \frac{i}{2}\ 
[G^{\alpha}(\xi),\Sigma(\xi)] + G^{\alpha}(\xi)\ F_i[\Sigma](\xi)\ .
\label{LIN4}
\end{equation}
Here $F_i[\Sigma](\xi)$ represents the internal gradient expansion in 
the self--energy $\Sigma$. Since the form (\ref{LIN4}) is fully general, 
it can be taken as a theorem:
\begin{itemize}
\item
{\it Internal} expansion of a term with a self--energy results by 
the substitution of $\Sigma(\xi)$ by $\Sigma(\xi) + F_i[\Sigma](\xi)$ 
in the zeroth order term.
\end{itemize}
From this theorem it follows that the {\it internal} gradient expansion 
in terms with a self--energy can be directly found from the function 
$F_i[\Sigma](\xi)$. 

%==============================================================================
\subsection{Expansions of a self--energy}
%==============================================================================
We can concentrate on this function $F_i[\Sigma](\xi)$ and evaluate 
it for several typical examples of a self--energy. 

%------------------------------------------------------------------------------
\subsubsection{Static averaged $T$--matrix approximation}
%------------------------------------------------------------------------------
The first example concerns the ATA self--energy (\ref{STM}) for electron 
scattering on local potentials. After application of the rules 1--3. 
the zeroth order propagator and correlation parts for the self--energy 
in a frequency representation results
\begin{equation}
\Sigma^r(\omega,T)=c\ \frac{V_0}{1-G^r(\omega,T)\ V_0}=c\ \theta^r(\omega,T)\ ,
\label{STr}
\end{equation}
\begin{equation}
\Sigma^<(\omega,T)=c\ \theta^r(\omega,T)\ G^<(\omega,T)\ \theta^a(\omega,T)
=c\ \theta^<(\omega,T)\ .
\label{ST<}
\vspace{2mm}
\end{equation}

Application of the rule 5. to the retarded part in (\ref{STMr}) gives the 
first order term 
\begin{equation}
F_i[\Sigma^r](\xi) =\frac{\delta^2 \Sigma^r}{\delta G^r \delta G^r}(\xi)\
\frac{i}{2}\ [G^r(\xi),G^r(\xi)] = 0\ ,
\label{FSTr}
\vspace{2mm}
\end{equation}
since only functional derivatives of $\Sigma^r$ over the propagator Green's 
functions $G^r$ can be applied here. Because the Poisson bracket from 
equivalent objects is zero, the function $F_i[\Sigma^r](\xi)$ does not 
contribute to the internal expansion.  Analogously can be found the 
internal term for the correlated part in (\ref{STM<})
$$
\vspace{2mm}
F_i[\Sigma^<](\xi)=
 \frac{\delta^2 \Sigma^<}{\delta G^r \delta G^<}(\xi)\ 
 \frac{i}{2}\ [G^r(\xi),G^<(\xi)]
+\frac{\delta^2 \Sigma^<}{\delta G^< \delta G^a}(\xi)\ 
\frac{i}{2}\ [G^<(\xi),G^a(\xi)]
$$
$$
+\frac{\delta^2 \Sigma^<}{\delta G^r \delta G^a}(\xi)\ 
\frac{i}{2}\ [G^r(\xi),G^a(\xi)]=...=
\vspace{3mm}
$$
$$
=-c\ {\rm Im \it} \left( (\theta^r(\omega,T))^2 \theta^a(\omega,T)\ 
[G^r(\omega,T), G^<(\omega,T)] \right)
-\frac{c}{2}\ |\theta^r(\omega,T)|^4\ G^<(\omega,T)\ 
\vspace{1mm}
$$
\begin{equation}
\times \int \frac{d\bar{\omega}}{2\pi}\ 
\frac{1}{\omega-\bar{\omega}}\ 
\left( \frac{\partial A(\bar{\omega},T)}{\partial \bar{\omega}} 
\frac{\partial A(\omega,T)}{\partial T}
-\frac{\partial A(\bar{\omega},T)}{\partial T} 
\frac{\partial A(\omega,T)}{\partial \omega} \right)\ .
\vspace{1mm}
\label{FST<}
\vspace{1mm}
\end{equation}
The term $F_i[\Sigma^>](\omega,T)$ can be evaluated in the same way.
In the first expression from (\ref{FST<}) only the nonzero functional 
derivatives have 
been considered, which result by differentiation over Green's functions of
different analytical structures. The order of derivatives and terms in 
Poisson brackets is the same as the order of these functions in (\ref{STM<}).
The second expression results by per--partes integration in the 
Poisson bracket $[G^r(\xi), G^a(\xi)]$ and some simple algebra.

%------------------------------------------------------------------------------
\subsubsection{Dynamic $T$--matrix approximation}
%------------------------------------------------------------------------------
Another $T$--matrix approximation is used \cite{KaBa,White}, if mutual 
interaction of electrons are studied. This approximation gives a 
self--energy of a very similar structure to the ATA self--energy in
(\ref{STM}), but internal dynamics is more complicated here.
For spinless Fermions the self--energy can be written as folows \cite{KaBa}
\begin{equation}
\vspace{2mm}
\Sigma(t_1,t_2) = -i\ \Theta(t_1,t_2)\ G(t_2,t_1)\ ,
\label{STh}
\vspace{2mm}
\end{equation}
where $\Theta$ is the dynamic $T$--matrix
$$
\vspace{2mm}
\Theta(t_1,t_2) = V\ \delta(t_1-t_2)+V\ R_0(t_1,\bar{t}_3)\ 
\Theta(\bar{t}_3,t_2)\ ,\ \ 
$$
\begin{equation}
R_0(t_1,t_2) = i\ G(t_1,t_2)\ G(t_1,t_2)\ .
\label{Th}
\vspace{2mm}
\end{equation}
In this self--energy the singular Hartree term is included ($t_2 
\rightarrow t_1^{-}$), but the exchange terms have been neglected 
for simplicity. 

Analytical continuation to real times of (\ref{Th}) gives the 
propagators and correlation functions (see the Appendix B)
$$
\vspace{2mm}
R_0^r(t_1,t_2)= G^r(t_1,t_2)\ G^>(t_1,t_2) 
               -G^<(t_1,t_2)\ G^r(t_1,t_2)\ ,
$$
\begin{equation}
\vspace{2mm}
R_0^<(t_1,t_2)= G^<(t_1,t_2)\ G^<(t_1,t_2)
\label{Ror<}
\end{equation}
and
$$
\vspace{2mm}
\Theta^r(t_1,t_2)=V\ \delta(t_1-t_2)+V\ R_0^r(t_1,\bar{t}_3)\ 
\Theta^r(\bar{t}_3,t_2)\ ,
$$
\begin{equation}
\vspace{2mm}
\Theta^<(t_1,t_2)= V\ [ \Theta^r(t_1,\bar{t}_3)\ R_0^<(\bar{t}_3,t_2) 
+                    \Theta^<(t_1,\bar{t}_3)\ R_0^a(\bar{t}_3,t_2)]\ .
\label{EThr<}
\vspace{2mm}
\end{equation}
After application of the rule 3. in (\ref{Ror<}--\ref{EThr<}) the zeroth 
order terms can be completed in a form similar to the $\theta$--functions 
in (\ref{STr}--\ref{ST<})
\begin{equation}
\vspace{2mm}
\Theta^r(\omega,T)=\frac{V}{1-R_0^r(\omega,T)\ V}\ ,\ \ \
\vspace{2mm}
\Theta^<(\omega,T)=\Theta^r(\omega,T)\ R_0^<(\omega,T)\ \Theta^a(\omega,T)\ .
\label{Thr<}
\end{equation}
The zeroth order contributions to the propagator and correlated parts 
of the self--energy (\ref{STh}) results from these functions as follows
$$
\vspace{2mm}
\Sigma^r(\omega,T)=
 \Theta^r(\omega+\bar{\omega},T)\ G^<(\bar{\omega},T)
-\Theta^<(\omega+\bar{\omega},T)\ G^a(\bar{\omega},T)\ ,
\vspace{2mm}
$$
\begin{equation}
\vspace{2mm}
\Sigma^<(\omega,T) = \Theta^<(\omega+\bar{\omega},T)\ G^>(\bar{\omega},T)\ .
\label{STDr<}
\vspace{2mm}
\end{equation}

We can continue with the internal gradient expansion in the self--energy
(\ref{STh}). The separate Green's function in (\ref{STDr<}) is on the 
highest level, so it is excluded from the internal expansion.  The 
remaining functions $\Theta^r$ and $\Theta^<$ could be independently 
expanded after the rule 4. and give $F_i[\Theta^r](\xi)$ and 
$F_i[\Theta^<](\xi)$.

Time arguments of the objects $R_0(t_1,t_2)$ in the function $\Theta(t_1,t_2)$ 
from (\ref{Th}) are in a series, so that the rule 5. can be directly 
applied. The expansion of the propagator function in terms of $R_0^r$ 
results zero, similarly as in (\ref{FSTr})
\begin{equation}
\vspace{2mm}
F_i[\Theta^r](\xi) =
\frac{\delta^2 \Theta^r}{\delta R_0^r \delta R_0^r}(\xi)\
\frac{i}{2}\ [R_0^r(\xi),R_0^r(\xi)] = 0\ .
\label{dThr}
\vspace{2mm}
\end{equation}
The internal expansions of the correlation function $\Theta^<$ is formed by
terms analogous those in (\ref{FST<})
$$
\vspace{2mm}
F_i[\Theta^<](\xi) =
  \frac{\delta^2 \Theta^<}{\delta R_0^r \delta R_0^<}(\xi)\
  \frac{i}{2}\ [R_0^r(\xi),R_0^<(\xi)]
+ \frac{\delta^2 \Theta^<}{\delta R_0^< \delta R_0^a}(\xi)\
  \frac{i}{2}\ [R_0^<(\xi),R_0^a(\xi)]
$$
$$
\vspace{2mm}
+ \frac{\delta^2 \Theta^<}{\delta R_0^r \delta R_0^a}(\xi)\
  \frac{i}{2}\ [R_0^r(\xi),R_0^a(\xi)] =...=
\vspace{2mm}
$$
$$
=-\ {\rm Im \it} \left( (\Theta^r(\omega,T))^2 \Theta^a(\omega,T)\ 
[R_0^r(\omega,T), R_0^<(\omega,T)] \right)
-2\ |\Theta^r(\omega,T)|^4\ R_0^<(\omega,T)\ 
\vspace{1mm}
$$
\begin{equation}
\times \int \frac{d\bar{\omega}}{2\pi}\ 
\frac{1}{\omega-\bar{\omega}}\ 
\left( \frac{\partial {\rm Im \mit} R_0^r (\bar{\omega},T)}{\partial \bar{\omega}} 
\frac{\partial {\rm Im \mit} R_0^r (\omega,T)}{\partial T}
-\frac{\partial {\rm Im \mit} R_0^r (\bar{\omega},T)}{\partial T} 
\frac{\partial {\rm Im \mit} R_0^r (\omega,T)}{\partial \omega} \right)\ .
\vspace{1mm}
\label{dTh<}
\end{equation}
\\
The function $F_i[\Theta^>(\xi)]$ results by the change of the index 
$<$ by $>$ in all places of (\ref{dTh<}).  Finally the internal 
expansions $F_i[\Sigma^r](\xi)$ and $F_i[\Sigma^<](\xi)$ can be 
obtained, if the functions $\Theta^r(\xi)$ and $\Theta^<(\xi)$ in the zeroth 
order self--energy (\ref{STDr<}) are substituted by $F_i[\Theta^r](\xi)$ 
and $F_i[\Theta^<](\xi)$ from (\ref{dThr}--\ref{dTh<}) 
$$
\vspace{2mm}
F_i[\Sigma^r](\omega,T) = F_i[\Theta^<](\omega+\bar{\omega},T)\ 
G^a(\bar{\omega},T)\ ,
$$
\begin{equation}
\vspace{2mm}
F_i[\Sigma^<](\omega,T) = F_i[\Theta^<](\omega+\bar{\omega},T)\ 
G^>(\bar{\omega},T)\ .
\label{FsSr<}
\vspace{2mm}
\end{equation}
Here the fact that $F_i[\Theta^r](\xi) = 0$ in (\ref{dThr}) has been
taken into account.

%------------------------------------------------------------------------------
\subsubsection{Shielded potential approximation}
%------------------------------------------------------------------------------
Particle interactions are often long range, like in the Coulomb potential.
If free particles are available in the system, then screening can
shorten the range of the interaction (give a much more localized potential). 
The shielded potential approximation of a self--energy \cite{KaBa} 
does not include ladders, like in the above studied examples, but rows of 
electron--hole bubbles screening the potential. Localization of the
potential by screening has a dynamical character. Therefore it would be
demanding to know time dependent behavior of screening in nonequilibrium 
processes.  Recently dynamics of ultrafast screening processes have been 
studied by NGF \cite{Hu,El-Sayed}. Close to equilibrium these systems can 
be described by the GBE, where the internal corrections to the screening 
dynamics can play an important role.

The shielded approximation for the self--energy can be written in the 
form \cite{KaBa} 
\begin{equation}
\vspace{2mm}
\Sigma(1,2) = i\ V_s(1,2)\ G(1,2)\ ,
\label{SSc}
\end{equation}
where the singular Fock term is included ($t_2 \rightarrow t_1^{+}$).
In the lowest order 
of a perturbation theory, the screened potential $V_s$ is related to 
the unscreened one $V$ by the polarization function $L_0$ as follows 
\vspace{2mm}
$$
V_s(1,2) = V(1,2)-V(1,\bar{3})\ L_0(\bar{3},\bar{4})\ V_s(\bar{4},2)\ ,\ \ \
V(1,2)=V(r_1-r_2)\ \delta(t_1-t_2)\ ,
$$
\begin{equation}
L_0(1,2) = i\ G(1,2)\ G(2,1)\ .
\label{VS}
\vspace{2mm}
\end{equation}

The propagators and correlation functions for the polarization function $L_0$
in (\ref{VS}) can be found by the rules in the Appendix B 
$$
L_0^r(1,2) = -(G^r(1,2)\ G^<(2,1) + G^<(2,1)\ G^a(2,1))\ ,\ \ \
$$
\begin{equation}
L_0^<(1,2) = -G^<(1,2)\ G^>(2,1)\ ,
\label{Lr<}
\vspace{2mm}
\end{equation}
and the functions $V_s^r$ and $V_s^<$ result analogously as $\Theta^r$
and $\Theta^<$ in (\ref{EThr<}).

Therefore after application of the rule 3. in (\ref{Lr<}), the zeroth order 
of the gradient expansion can be obtained as in (\ref{Thr<}) (the full CMS
coordinates $\xi=(k,\omega;R,T)$ are considered, since the interaction is 
nonlocal)
\begin{equation}
\vspace{2mm}
V_s^r(\xi)=\frac{V(k)}{1+L_0^r(\xi)\ V(k)}\ ,\ \ \
V_s^<(\xi)=V_s^r(\xi)\ L_0^<(\xi)\ V_s^a(\xi)\ .
\label{VSr<}
\vspace{2mm}
\end{equation}
Similarly the zeroth order gradient contributions to the propagator and 
correlated parts of the self--energy (\ref{SSc}) result
$$
\Sigma^r(k,\omega;R,T)=
 G^r(k-\bar{k},\omega-\bar{\omega};R,T)\ V_s^>(\bar{k},\bar{\omega};R,T)
$$
$$
-G^<(k-\bar{k},\omega-\bar{\omega};R,T)\ V_s^a(\bar{k},\bar{\omega};R,T)\ ,
\vspace{5mm}
$$
\begin{equation}
\Sigma^<(k,\omega;R,T) = G^<(k-\bar{k},\omega-\bar{\omega};R,T)\
V_s^<(\bar{k},\bar{\omega};R,T)\ .
\label{SScr<}
\vspace{2mm}
\end{equation}

The structure of the internal corrections to the self--energy (\ref{VS}) 
is also similar to the previous example.  The main difference is in that 
here the potential $V(k)$ depends on the wave vector $k$, so that
new corrections terms are included
$$
\vspace{2mm}
F_i[V_s^r](\xi) =
\frac{\delta^2 V_s^r}{\delta L_0^r \delta L_0^r}(\xi)\
\frac{i}{2}\ [L_0^r(\xi),L_0^r(\xi)] 
\vspace{2mm}
$$
\begin{equation}
+\frac{\delta^2 V_s^r}{\delta L_0^r \delta V}(\xi)\
\frac{i}{2}\ [L_0^r(\xi),V(k)] 
+\frac{\delta^2 V_s^r}{\delta V \delta L_0^r}(\xi)\
\frac{i}{2}\ [V(k),L_0^r(\xi)] = 0\ .
\label{dVSr}
\end{equation}
\\
The potential $V$ in the last two terms is either more 'right'
or more 'left' than $L_0$ in the series (\ref{VS}) (see also 
comment below (\ref{FST<})).  Since the second functional derivative 
is symmetrical in the differentiating functions, these two terms 
diminish each other.
In the Poisson brackets from (\ref{dVSr}) derivatives over 
space--momentum coordinates are also performed (see (\ref{SLOW2})), but
the potential $V(k)$ can be differentiated only over the momentum.

The correlation function $V_s^<$ gives more interesting internal 
contributions.  We write only terms with nonzero Poisson brackets and 
also neglect terms analogous those in (\ref{dVSr}), where both 
functions originate from one of the propagators $V_s^{r,a}$. 
The fact that the differentiating functions $V$, $L_0^{r,a}$ 
originate from different full potential propagators $V_s^{r,a}$ 
is symbolized by a small index at the potentials $V^{(r)},\ V^{(a)}$, 
which show the side they originate from. The nonzero terms are
$$
\vspace{3mm}
F_i[V_s^<](\xi) =
  \frac{\delta^2 V_s^<}{\delta L_0^r \delta L_0^<}(\xi)\
  \frac{i}{2}\ [L_0^r(\xi),L_0^<(\xi)]
+ \frac{\delta^2 V_s^<}{\delta L_0^< \delta L_0^a}(\xi)\
  \frac{i}{2}\ [L_0^<(\xi),L_0^a(\xi)]
$$
$$
+ \frac{\delta^2 V_s^<}{\delta L_0^r \delta L_0^a}(\xi)\
  \frac{i}{2}\ [L_0^r(\xi),L_0^a(\xi)]\ 
$$
$$
+ \frac{\delta^2 V_s^<}{\delta L_0^r \delta V^{(a)}}(\xi)\
  \frac{i}{2}\ [L_0^r(\xi),V(k)]
+ \frac{\delta^2 V_s^<}{\delta V^{(r)} \delta L_0^a}(\xi)\
  \frac{i}{2}\ [V(k),L_0^a(\xi)]
\vspace{3mm}
$$
\begin{equation}
+ \frac{\delta^2 V_s^<}{\delta L_0^< \delta V^{(a)}}(\xi)\
  \frac{i}{2}\ [L_0^<(\xi),V(k)]
+ \frac{\delta^2 V_s^<}{\delta V^{(r)} \delta L_0^<}(\xi)\
  \frac{i}{2}\ [V(k),L_0^<(\xi)]\ .
\label{dVS<1}
\end{equation}
\\
The first three terms in the right side of (\ref{dVS<1}) are analogous 
those in (\ref{dTh<}), but the additional space--momentum derivatives 
present in the Poisson brackets here. The next two terms in (\ref{dVS<1})
are evidently nonzero, because the derivatives and the Poisson brackets 
are different and nonzero. In the remaining two terms the Poisson 
brackets are opposite each other, but the functional derivatives are 
different, because the potentials are taken from different sides.

We can evaluate the above functional derivatives and collect all the terms 
$$
F_i[V_s^<](\xi) 
= {\rm Im \it}\left\{ (V_s^r(\xi))^2\ V_s^a(\xi)\
\left( [L_0^r(\xi),L_0^<(\xi)]
- \frac{1}{V(k)^2}\ [V(k),L_0^<(\xi)] \right) \right\}
$$
\begin{equation}
+ (V_s^r(\xi))^2\ L_0^<(\xi)\ (V_s^a(\xi))^2\
\frac{i}{2}\ \left( [L_0^r(\xi),L_0^a(\xi)]\ 
+ \frac{1}{V(k)^2}\ [V(k),L_0^r(\xi)-L_0^a(\xi)] \right)\ .
\label{dVS<2}
\vspace{2mm}
\end{equation}
The function $F_i[V_s^>(\xi)]$ results by the change of the index $<$ by $>$
in all places of (\ref{dVS<2}).

As before the internal expansions $F_i[\Sigma^r](\xi)$ and 
$F_i[\Sigma^<](\xi)$ can 
be obtained, if the functions $V_s^{r,a}(\xi)$ and $V_s^{<,>}(\xi)$ in 
the zeroth order self--energy (\ref{SScr<}) are substituted by 
$F_i[V_s^{r,a}](\xi)$ and $F_i[V_s^{<,>}](\xi)$ from (\ref{dVSr}--\ref{dVS<2}) 
$$
\vspace{2mm}
F_i[\Sigma^r] (k,\omega;R,T) =G^r(k-\bar{k},\omega-\bar{\omega};R,T)\ 
F_i[V_s^>](\bar{k},\bar{\omega};R,T)\ ,
$$
\begin{equation}
\vspace{2mm}
F_i[\Sigma^<] (k,\omega;R,T) = G^<(k-\bar{k},\omega-\bar{\omega};R,T)\ 
F_i[V_s^<](\bar{k},\bar{\omega};R,T)\ .
\label{FsScr<}
\vspace{2mm}
\end{equation}
The fact that $F_i[V_s^{r,a}](\xi) = 0$ in (\ref{dVSr}) has been
again taken into account.

\newpage
%%%%%%%%%%%%%%%%%%%%%%%%%%%%%%%%%%%%%%%%%%%%%%%%%%%%%%%%%%%%%%%%%%%%%%%%%%%%%%%
\section{Transport equations with internal corrections}
%%%%%%%%%%%%%%%%%%%%%%%%%%%%%%%%%%%%%%%%%%%%%%%%%%%%%%%%%%%%%%%%%%%%%%%%%%%%%%%
The internal corrections might appear in all kind of equations derived 
by the gradient expansions from the Kadanoff--Baym equations.

%==============================================================================
\subsection{Generalized Boltzmann equation}
%==============================================================================
The complete generalized Boltzman equation with both the {\it external} 
and {\it internal} gradient corrections can be found directly, when the 
above theorem is implemented in its derivation from the Appendix B. 
The theorem says that the internal terms result by the substitution of
$\Sigma^{<(>)}(\xi)$ by $\Sigma^{<(>)}(\xi)+ F_i[\Sigma^{<(>)}](\xi)$ in
the zeroth order terms. Therefore after a Fourier transform the complete 
GBE gets the form
\vspace{2mm}
$$
\left(\frac{\partial }{\partial T}+ \frac{{\bf p} \cdot \nabla_R}{m} 
-\nabla_R U_{eff}(\xi) \cdot \nabla_{\bf p} 
+\frac{\partial U_{eff}(\xi)}{\partial T}\ 
\frac{\partial }{\partial \omega} \right)\ G^<(\xi)
$$
$$
- [ {\rm Re \it}\Sigma^r(\xi),G^<(\xi)]
+ [ {\rm Re \it}G^r(\xi),\Sigma^<(\xi)]
\vspace{3mm}
$$
\begin{equation}
=-\left(\Sigma^>(\xi)+F_i[\Sigma^>(\xi)]\right) G^<(\xi)
 +\left(\Sigma^<(\xi)+F_i[\Sigma^<(\xi)]\right) G^>(\xi)\ .
\label{GD<}
\vspace{3mm}
\end{equation}
The equation for the correlation function $G^>(\xi)$ can be found similarly.
The singular (Hartree--Fock) contributions in (\ref{GD<}) are included 
in the effective potential $U_{eff}(\xi)$.
The propagator functions $G^{r(a)}(\xi)$ and $\Sigma^{r(a)}(\xi)$ in
(\ref{GD<}) are related to the correlated parts $G^<(\xi)$, $G^>(\xi)$
and $\Sigma^<(\xi)$, $\Sigma^>(\xi)$ by the Hilbert transform (\ref{Selfr}). 
Therefore it is not necessary to find separate transport equations for 
these propagator functions, which would eventually include the nonzero 
internal terms $F_i[\Sigma^{r(a)}(\xi)]$. The {\it internal} terms in
(\ref{GD<}) contribute to the dynamics (not the renormalization) of the 
studied system, so that they cannot be neglected in the transport equations. 
Conservation laws for the GBE in (\ref{GD<}) could be proven similarly as
for the exact Kadanoff--Baym equations \cite{KaBa}.

In the above example of the ATA self--energy, the expressions 
$F_i[\Sigma^{<(>)}(\xi)]$ in (\ref{GD<}) should be substituted by 
(\ref{FST<}). For the dynamic $T$--matrix approximation the terms
(\ref{FsSr<}) fulfill this role (the Hartree term should be taken once).
In both these examples the CMS coordinates in the functions 
$F_i[\Sigma^{<(>)}(\omega,T)]$ are reduced with respect to the other terms in 
(\ref{GD<}), where it is $\xi=(k,\omega;R,T)$. This reduction results from 
the local form of the $T$--matrix approximations studied here. In the 
shielded potential approximation for the self--energy, the expressions 
$F_i[\Sigma^{<(>)}(\xi)]$ from (\ref{FsScr<}) with full CMS coordinates 
can be used in the GBE. In all these cases the complete GBE with internal 
correction terms results quite complicated. Nevertheless, we believe 
that it can be handled by some approximate numerical methods.

%==============================================================================
\subsection{Linearized transport equations}
%==============================================================================
The internal gradient corrections can appear also in linearized 
transport equations derived from the GBE in weak $dc$--electric fields 
\cite{Hansch}. It has been already mentioned \cite{Chen} that these 
linearized equations should include further correction terms for 
complicated scattering. 

The left side of the GBE in (\ref{GD<}) looks in $dc$--electric fields 
as follows 
\begin{equation}
\left(\frac{\partial }{\partial T}+ \frac{1}{m}({\bf k} +e T {\bf E_0}) \cdot 
\left( \nabla_R +e {\bf E_0}\frac{\partial}{\partial \omega}\right) \right)\ 
G^<(\xi)\ .
\label{drive}
\end{equation}
To diminish the explicit time dependence in (\ref{drive}), not present in 
dissipative systems in weak $dc$--fields, the following transform 
should be performed \cite{Hansch}
\begin{equation}
{\bf Q} \rightarrow {\bf k}+e {\bf E_0} T\ ,\ \
\frac{\partial}{\partial T} \rightarrow \frac{\partial }{\partial T} 
+e{\bf E_0}\cdot \nabla_{\bf Q}\ , 
\label{tran}
\end{equation}
where ${\bf E_0}$ is the intensity of the $dc$--electric field.
Application of this transform to the Poisson brackets in the second
line of (\ref{GD<}) gives mixed terms in the following way
\begin{equation}
[A,B] \rightarrow [A,B]+e {\bf E_0}\cdot 
\left( \frac{ \partial A}{\partial \omega}\ \nabla_{\bf Q} B
- \nabla_{\bf Q} A\ \frac{ \partial B}{\partial \omega}\ \right)\ .
\label{term}
\end{equation}
If this transform is applied also to the Poisson brackets in the internal 
corrections $F_i[\Sigma^{<(>)}(\xi)]$ in (\ref{GD<}), further new terms can
result.

When the self--energy $\Sigma(\xi)$ depends only on $(\omega,T)$ variables,
like in the space localized scattering (\ref{STM}) or (\ref{STh}), 
then only the external terms $\frac{\partial {\rm Re \it}\Sigma^r(\omega)}
{\partial \omega} \frac{\partial G^{<,>}(k,\omega)}{\partial k}$ and
$\frac{\partial \Sigma^{<,>}(\omega)} {\partial \omega}
\frac{\partial  {\rm Re \it}G^r(k,\omega)}{\partial k}$,
result by the transform (\ref{term}). Application of this transform
in the terms $F_i[\Sigma^{<(>)}(\xi)]$ from (\ref{FST<}), (\ref{FsSr<})
give nothing, because only $k$--independent Green's functions are present
there.  If a nonlocal ATA self--energy \cite{JauWil} is used, generalizing 
(for neutral smeared imperfections) the local form (\ref{STM}), then 
new terms would result from application of the transform (\ref{term}) 
in the $k$--dependent internal corrections $F_i[\Sigma^{<(>)}(\xi)]$.
The nonlocal scattering results for example also on charged impurities 
\cite{Grill}, where it is reasonable to screen the impurity potential 
\cite{HuSar}, similarly as in the self--energy (\ref{SSc}). 
In both these examples internal and external corrections in (\ref{GD<})
are nonzero.  Since evaluation of these terms is direct, we do 
not write them as well as the resulting complicated linearized equations.  

The question is how the internal corrections can contribute in the
case of the linear response to weak $ac$--electric fields, where we would
expect that also the space--local interactions (\ref{STM}), (\ref{STh}) 
give new correction terms. The linearized transport equations in the
$ac$--electric fields  have been studied \cite{Wu}, but the transform 
to new coordinates \cite{Hansch} was
not performed. Therefore no gradient corrections would seemingly
contribute by new terms. In fact it is probably hard or not unique 
to find the above transform in the $ac$--case. Moreover in the 
$ac$--case it is not sufficient to stop the gradient expansions 
in the lowest orders. Therefore a new gauge invariant approach 
has been developed \cite{Levanda} to study linear response to the
weak $ac$ and $dc$--electric fields, where no additional transforms are 
necessary. Unfortunately, the resulting equations are still quite 
complicated. Recently a relatively simple consistent approach has been 
devised \cite{Kral}, which starts from the integral version of the 
Kadanoff--Baym equations.

%%%%%%%%%%%%%%%%%%%%%%%%%%%%%%%%%%%%%%%%%%%%%%%%%%%%%%%%%%%%%%%%%%%%%%%%%%%%%%%
\section{Conclusion}
%%%%%%%%%%%%%%%%%%%%%%%%%%%%%%%%%%%%%%%%%%%%%%%%%%%%%%%%%%%%%%%%%%%%%%%%%%%%%%%
We have found new gradient corrections in the generalized Boltzmann 
equation \cite{KaBa}. These corrections result if the gradient expansion 
is performed also inside of the self--energy in scattering integrals 
of the quantum transport equations. We call these corrections {\it 
internal}, because they reflect the many--body character of scattering 
processes represented by the internal structure of the self--energy. 
Analogously the standard gradient corrections \cite{KaBa}, which do not 
take into account the internal structure of the self--energy, are called here 
{\it external}.

The generalized Boltzmann equation with all correction terms has been
derived. The internal corrections to the GBE have been calculated 
for electron scattering on localized static potentials, which is 
described by the ATA self--energy. More complex corrections have been
obtained for interacting spinless Fermions, where the self--energy is 
described either by a local $T$--matrix approximation or by a nonlocal 
shielded potential approximation. We believe that the GBE with the new
correction terms might be a proper tool for studies of relaxation
to equilibrium in systems with nontrivial electron interactions. We 
are planing to investigate some of these systems in future.

We have also discussed the importance of the internal corrections in
the linearized transport equations in weak electric fields \cite{Hansch},
which can be derived from the GBE. The presence of new correction terms 
in the $dc$--version of these equations has been clarified in the 
above studied examples.  The internal corrections might be important 
in many other physical problems, where the self--energy includes multiple 
scattering events.

\vspace{5mm}
\noindent {\bf Acknowledgment}

\noindent The author would like to thank B. Velick\'y and V. \v C\' apek 
for valuable comments.

\newpage
%...........................................
\vspace{7mm}
{\bf \Large Appendix A}
\vspace{3mm}
\setcounter{equation}{0}
\renewcommand{\theequation}{A.\arabic{equation}}
%...........................................

\noindent
The causal Fermion ($O=\psi$) or Boson ($O=A$) Green's functions in real
times are defined by \cite{Abrikosov,Mahanb} (Matsubara Green's functions
in complex times look analogously)
\vspace{1mm}
\begin{equation}
{G^t(1,2)=-\frac{i}{\hbar}<T[O(1)\ O^{\dagger}(2)]>\ ,\ \
j \equiv (r_j,t_j)\ ,\ \  } (j=1,\ 2)\ .
\label{Gc}
\vspace{1mm}
\end{equation}
Correlation functions are related to the causal function as follows
\vspace{1mm}
$$
    i\hbar\ G^t(1,2)=G^>(1,2)=<O(1)\ O^{\dagger}(2)>\ ,\ \ t_1>t_2\ ,
$$
\begin{equation}
\mp i\hbar\ G^t(1,2)=G^<(1,2)=<O^{\dagger}(2)\ O(1)>\ ,\ \ t_1<t_2\ ,
\label{G<>}
\vspace{1mm}
\end{equation}
where the upper (lower) sign applies to Fermions (Bosons).

The retarded and advanced Green's functions are defined by
\vspace{1mm}
$$
G^r(1,2)= -\frac{i}{\hbar}\ \theta(1-2)\ [ G^>(1,2) \pm G^<(1,2) ]\ ,
$$
\begin{equation}
G^a(1,2)=  \frac{i}{\hbar}\ \theta(2-1)\ [ G^>(1,2) \pm G^<(1,2) ]\ ,
\label{Gra}
\vspace{1mm}
\end{equation}
where the theta function is $\theta(t)=0,\ t<0; \ \theta(t)=1,\ t\geq 0$. 

In equilibrium and space homogeneous systems the Green's functions 
depend only on the difference of coordinates $(r,t)= (r_1- r_2, t_1-t_2)$, 
so that they can be easily Fourier transformed to the 
$(k,\omega)$--representation as follows
\vspace{1mm}
\begin{equation}
G(k,\omega) = \int d^n r \int dt\
\exp(i(\omega t - r\cdot k))\ G(r_1-r_2;t_1-t_2) \ .
\label{Gck}
\vspace{1mm}
\end{equation}
Then the Fermion and Boson correlation functions can be expressed as 
\cite{KaBa}
\vspace{1mm}
\begin{equation}
G^<(k,\omega)=n_{F,B}(\hbar\omega)\ A(k,\omega)\ ,\ \
G^>(k,\omega)=(1 \mp n_{F,B}(\hbar\omega))\ A(k,\omega)\ ,
\label{cor1}
\vspace{1mm}
\end{equation}
where $n_F,\ n_B$ denote the Fermi--Dirac and Bose--Einstein distributions
$$
n_{F,B}(\hbar\omega)=\frac{1}{e^{\frac{\hbar\omega}{kT}} \pm 1}\
$$
and the spectral function is defined by
\vspace{1mm}
\begin{equation}
A(k,\omega) \equiv -2\ {\rm Im \mit}\ G^r(k,\omega)
=G^>(k,\omega) \pm G^<(k,\omega)\ .
\label{cor2}
\vspace{1mm}
\end{equation}

The retarded Green's function can be calculated from the spectral 
function (\ref{cor2}) as follows ((\ref{cor2}--\ref{Selfr}) hold
also in full CMS coordinates $\xi=(k,\omega;R,T)$)
\vspace{1mm}
\begin{equation}
G^r(k,\omega)= \int_{-\infty}^{\infty}\ \frac{d\bar{\omega}}{2\pi}\
\frac{A(k,\bar{\omega})}{\omega-\bar{\omega}+i\delta}\ .
\label{Selfr}
\end{equation}
Similar formulas can be applied also for the self--energy. 

%...........................................
\vspace{7mm}
{\bf \Large Appendix B}
\vspace{3mm}
\setcounter{equation}{0}
\renewcommand{\theequation}{B.\arabic{equation}}
%...........................................

\noindent
The nonequilibrium Green's functions can be found by analytical
continuation to real times of the Matsubara Green's functions in 
complex times \cite{KaBa}. In NGF it is often necessary to find 
the propagator or correlation  part of a combination of functions.
An example is the product
\begin{equation}
A(1,2)=B(1,2)\ C(1,2)\ ,
\label{ABC}
\end{equation}
where $A,\ B,\ C$ are one-particle causal Green's functions or
self--energies. The required functions can be found by LW rules 
\cite{Langreth}, where the signs and prefactors result from the definitions 
(\ref{G<>}--\ref{Gra}). We have found the expressions for (\ref{ABC})
in two cases, where the functions $A,\ B,\ C$ correspond either to Fermions 
(F) or to Bosons (B) as follows: 
(1) $(A,B,C)=(F,F,B)$ (an example is an electron--phonon self--energy 
\cite{Mahanb}) or
(2) $(A,B,C)=(B,F,F)$ (an example is the function $R_0$ in (\ref{Th})).
In both these cases the expressions result in the form
$$
A^<(1,2)=-i\ B^<(1,2)\ C^<(1,2)\ ,\ \ \
A^>(1,2)=-i\ B^>(1,2)\ C^>(1,2)\ ,
$$
$$
A^r(1,2)=-i\ (B^r(1,2)\ C^>(1,2) - B^<(1,2)\ C^r(1,2) )\ ,
$$
\begin{equation}
A^a(1,2)=-i\ (B^a(1,2)\ C^>(1,2) - B^<(1,2)\ C^a(1,2) )\ .
\label{Gr<}
\vspace{1mm}
\end{equation}
Similarly can be found the propagators and correlation functions for
the expression 
\begin{equation}
A(1,2)=B(1,2)\ C(2,1)\ ,
\label{ABCi}
\vspace{1mm}
\end{equation}
where the following possibilities have been chosen:
(1) $(A,B,C)=(F,B,F)$ (an example is the self--energy (\ref{STh})) or
(2) $(A,B,C)=(B,F,F)$ (an example is the electron--hole bubble $L_0$ 
in (\ref{VS})).
These possibilities, which, but the order of $B,\ F$ in (1), are the same 
as in (\ref{Gr<}), give the identities ((1) for ($-$), (2) for ($+$))
$$
A^<(1,2)=i\ B^<(1,2)\ C^>(2,1)\ ,\ \ \
A^>(1,2)=i\ B^>(1,2)\ C^<(2,1)\ ,
$$
$$
A^r(1,2)=i\ (B^r(1,2)\ C^<(2,1) \mp B^<(1,2)\ C^a(2,1))\ ,
$$
\begin{equation}
A^a(1,2)=i\ (B^a(1,2)\ C^<(2,1) \mp B^<(1,2)\ C^r(2,1))\ .
\label{Gr<i}
\vspace{1mm}
\end{equation}
The following structure appears also in most formulas
\begin{equation}
A(1,2)=B(1,\bar{3})\ C(\bar{3},2)\ ,
\label{AB-C}
\vspace{1mm}
\end{equation}
where the bars over the arguments mean integration over the whole real 
axis of these coordinates. The expressions result the same, 
irrespective of the types of involved functions
$$
A^<(1,2)= B^r(1,\bar{3})\ C^<(\bar{3},2)+ B^<(1,\bar{3})\ C^a(\bar{3},2)\ ,
$$
$$
A^>(1,2)= B^r(1,\bar{3})\ C^>(\bar{3},2)+ B^>(1,\bar{3})\ C^a(\bar{3},2)\ ,
\vspace{2mm}
$$
\begin{equation}
A^r(1,2)=B^r(1,\bar{3})\ C^r(\bar{3},2)\ , \ \ \
A^a(1,2)=B^a(1,\bar{3})\ C^a(\bar{3},2)\ .
\label{Gr<-}
\vspace{1mm}
\end{equation}
In the text we use the term 'parallel' for the structures of
arguments in (\ref{ABC}--\ref{Gr<i}), while the structure 
(\ref{AB-C}--\ref{Gr<-}) is termed 'serial'.

The Kadanoff--Baym equations can be found by application of the rules 
(\ref{Gr<-}) to the differential Dyson equation, which can be written
in two forms. If we take into account that $(G_0^{-1})^<(1,2)=0$ and 
$\delta^<(1-2)=0$, then the two forms of the Kadanoff--Baym equations 
look as follows
$$
(G_0^r)^{-1}(1,\bar{3})\ G^<(\bar{3},2)
=\Sigma^r(1,\bar{3})\ G^<(\bar{3},2)+\Sigma^<(1,\bar{3})\ G^a(\bar{3},2)\ ,
$$
\begin{equation}
G^<(1,\bar{3})\ (G_0^a)^{-1}(\bar{3},2)
=G^r(1,\bar{3})\ \Sigma^<(\bar{3},2)+G^<(1,\bar{3})\ \Sigma^a(\bar{3},2)\ .
\label{KBE}
\end{equation}
Analogous equations can be obtained for the correlation function $G^>$.

The generalized Boltzmann equation \cite{KaBa} can be derived by 
subtraction of the two sets of equations (\ref{KBE}). In the resulting 
quantum transport equations for $G^{<,>}$ it is necessary to introduce 
the CMS coordinates and perform the gradient expansion up to
the first order. Then a Fourier transform
over the small coordinates is performed. It is helpful to resolve the 
propagators $G^{r,a}$, $\Sigma^{r,a}$ from the right side of (\ref{KBE}) into
the real and imaginary parts. The imaginary parts of these propagators
can be resolved with the help of the identity (\ref{cor2}) (in the CMS 
coordinates).  Then the terms with equal correlation signs $<,<$ and
$>,>$ fall out from the scattering side of the new equations and the
GBE easily results (see its completed form in (\ref{GD<})).

\end{document}